\documentclass[aps,twocolumn,showpacs]{revtex4}

\usepackage{graphicx}

\begin{document}
\title{Diffraction phases in atom interferometers}

\author{M. B\"uchner, R. Delhuille, A. Miffre, C. Robilliard, J.
Vigu\'e (a)
\\  and C. Champenois (b)}
\affiliation{(a)  Laboratoire Collisions Agr\'egats R\'eactivit\'e
-IRSAMC
\\ Universit\'e Paul Sabatier and CNRS UMR 5589, 118, Route de Narbonne 31062 Toulouse Cedex, France
\\(b) PIIM, Universit\'e de Provence and CNRS UMR 6633, Centre de saint J\'er\^ome case C21, 13397 Marseille cedex 20, France.
\\ e-mail:~{\tt jacques.vigue@irsamc.ups-tlse.fr}}



\begin{abstract}
Diffraction of atoms by laser is a very important tool for matter
wave optics. Although this process is well understood, the phase
shifts induced by this diffraction process are not well known. In
this paper, we make analytic calculations of these phase shifts in
some simple cases and we use these results to model the contrast
interferometer recently built by the group of D. Pritchard at MIT.
We thus show that the values of the diffraction phases are large
and that they probably contribute to the phase noise observed in
this experiment.

\end{abstract}


\maketitle


\section{Introduction}

In atom interferometry, laser diffraction is a very powerful and
versatile tool (for overviews, see references
\cite{berman97,crasc01}). The diffraction of matter waves by a
standing light wave was proposed by P. Kapitza and P.A.M. Dirac
\cite{kapitza33} in the case of electrons and generalized to atoms
by S. Altshuler et al. \cite{altshuler66}. Atom diffraction by
light has been studied theoretically \cite{cook78,bernhardt81} and
experimentally \cite{arimondo79,pritchard83} and these early works
have been followed by many studies too numerous to be quoted here.
The phases of the diffraction amplitudes are rarely discussed in
detail, with a few exceptions like the works of S. Chu and
coworkers \cite{weitz94} and of K. Burnett and coworkers
\cite{featonby96}, in both cases for Raman adiabatic transfer, and
the work of C. Bord\'e and coworkers \cite{borde97,borde99}, which
analyzes the general diffraction process in the rotating wave
approximation. Unfortunately, this approximation cannot be used
for elastic diffraction studied here.

In an interferometer, the diffraction phases modify the
interference signals but this effect is difficult to detect, as it
requires accurate phase measurements and it cancels in symmetric
interferometers, like the Mach-Zehnder interferometer. The goal of
this paper is to present an analytic calculation of diffraction
phases in a simple case (elastic diffraction by a laser standing
wave) and to show the importance of these diffraction phases in an
existing experiment. We consider here diffraction in the
Raman-Nath regime and second order Bragg diffraction in the weak
field regime and we apply these results to the contrast
interferometer built by the group of D. Pritchard \cite{gupta02}.
The calculated diffraction phases are large in this interferometer
and as these phases depend rapidly on the laser power density used
for the diffraction process, our calculation may explain the
observed phase noise as resulting from fluctuations of this power
density.

\section{The problem}

We consider diffraction of slow ground state atoms by a
near-resonant laser standing wave of frequency $\omega_L$. For a
sufficiently large laser detuning $\delta = \omega_L - \omega_0$,
where $\omega_0$ is the resonance transition frequency, the
probability of real excitation is negligible and the diffraction
process is coherent. In the dressed-atom picture
\cite{cohentannoudji88}, the laser standing wave creates a light
shift potential $V(x,t)$ :
\begin{eqnarray}
\label{potential}
 V(x,t) & = & V_0(t) \cos^2 (k_Lx)  \nonumber
\\ & = &  \frac{V_0(t)}{4} \left[ 2 +
  \exp(+2ik_Lx) + \exp(-2ik_Lx) \right]
\end{eqnarray} where the envelope $V_0(t)$ is  proportional to the
laser power density divided by the frequency detuning $\delta$ and
$k_L$ is the laser wavevector. We are going to forget the
$x$-independent term, which simply shifts the energy zero and
therefore has no effect, as long as all atoms see the same
potential. The motion along the $y$ and $z$ directions is free and
will not be discussed. The natural energy unit is the atom recoil
energy $ \hbar\omega_{rec} = \hbar^2 k_L^2/2m$ and we will measure
the potential with this unit, by defining $q(t)$
\cite{keller99,champenois01} :
\begin{equation}
\label{RU1}
 q(t) =V_0(t) / (4 \hbar\omega_{rec})
\end{equation} Using a dimensionless time
$\tau$ defined by $\tau =\omega_{rec}t$, a dimensionless spatial
coordinate, $X = k_L x$ and a dimensionless wavevector $\kappa =
k_x/k_L$, the 1D Schr\"odinger equation becomes :
\begin{equation}
\label{RU2} i\frac{\partial\Psi}{\partial \tau} = -
\frac{\partial^2\Psi}{\partial X^2} + q(\tau)\left[\exp(2i X) +
\exp(-2iX)\right] \Psi
\end{equation}

\begin{figure}[htb]
\includegraphics{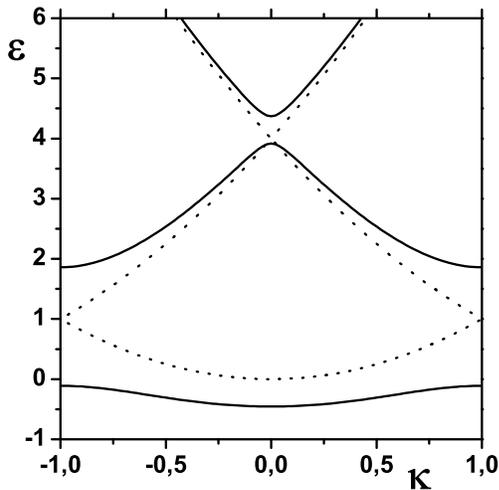}
\caption{\label{bloch energies} Plots of the energies
$\varepsilon$ of the lowest Bloch states versus the pseudomomentum
$\kappa$ : solid line $q= 1$; dashed line $q=0$.}
\end{figure}

For a constant value of the potential $q$, the atom eigenstates
are Bloch states \cite{horne99,keller99,champenois01}. Writing the
Hamiltonian matrix corresponding to equation (\ref{RU2}) in the
basis $\left|\kappa \right>$ of plane waves of momentum $\hbar
\kappa$ and using numerical diagonalization,  we get the band
structure $\varepsilon(\kappa,p)$, with the pseudo-momentum
$\kappa$ belonging to the first Brillouin zone ( $- 1< \kappa \leq
1$) and the integer $p$ labeling the bands \cite{champenois01}.
Figure~\ref{bloch energies} presents the energy of the lowest
Bloch states as a function of $\kappa$ for two values of the
potential, $q=0$ and $q=1$, with two important features : when $q$
is not equal to zero, band gaps appear at each crossing of the
$q=0$ folded parabola and energy shifts appear at the same time.
These energy shifts are explained by perturbation theory : each
free plane wave $\left|\kappa\right> $ is coupled to two other
states, $\left| \kappa \pm 2 \right>$ and the two coupling terms
are equal. As the energy denominator is larger for the coupling to
the upper state, all the levels are pushed upwards (except near
the places where gaps open), but the lowest Bloch state is
obviously pushed downwards.

\section{Diffraction phases}

In order to simplify the calculations, we consider that the atom
is initially in a state of zero momentum, $\left|\psi(\tau
=0)\right> = \left|0\right>$. We first consider diffraction in the
Raman-Nath regime. This approximation consists in neglecting the
dynamics of the atom during the diffraction process produced by a
pulse $q(\tau)$ of duration $\tau_{RN}$. This approximation is
good if the potential $q(\tau)$ is intense, $q\gg 1$, and if the
pulse is brief, $\tau_{RN}\ll 1$. The validity range of this
approximation is given by \cite{champenois01,keller99} :
\begin{equation}
\label{RN1}
 \tau_{RN}  < 1/(4\sqrt{q})
\end{equation}
\noindent and the diffracted wave is a classic result :
\begin{equation}
\label{RN2} \left|\psi(\tau_{RN})\right> = \sum_{p}  (-i)^{|p|}
J_{|p|}(\gamma) \left|2p\right>
\end{equation} with $\gamma = 2 q \tau_{RN}$. We have verified
\cite{champenois01} that the Raman-Nath formula predicts
accurately the diffraction probability of order $0$ and $1$, for
finite values of the parameter $q$, as long as condition
(\ref{RN1}) is verified, but we have not tested the phases of
these diffraction amplitudes. They could be tested by using the
diffraction amplitudes calculated \cite{berry66} as a power series
of $1/q$.

Second order Bragg diffraction is due to the indirect coupling of
the $\left|\pm2\right>$ free states, through the $\left|0\right>$
state. As this coupling is a second order term in $q$, to make a
consistent treatment, we must consider the 5 lowest energy states,
with $\kappa= 0, \pm2, \pm4$. The Hamiltonian matrix has the
following non-vanishing elements $ \left<2p|H|2p\right> = 4p^2$
and $ \left<2p|H|2(p\pm1)\right> = q$. Up to second order in $q$,
the energy correction of the $|0>$ state is $ E_0 = - q^2/2$ and
the effective Hamiltonian coupling the states $\left|-2\right>$
and $\left|+2\right>$ is:
\begin{equation}
\label{PR4} H_{eff}= \left[ {4 + (q^2/6) \atop (q^2/4)
}\quad{(q^2/4) \atop 4 + (q^2/6)}\right]
\end{equation}
We have tested the quality of this expansion limited to the $q^2$
terms, by numerical diagonalization of the Hamiltonian matrix. The
neglected terms (in $q^4$, etc.) are of the order of $1\%$
($10\%$) of the $q^2$ terms if $q=0.3$ ($q=1$ respectively), thus
giving an idea of the validity range of this calculation.

The dynamics is adiabatic if the potential $q(\tau)$ varies
slowly, but diffraction remains possible when two free states are
degenerate, as the $\left|\pm2\right>$ states. The problem is
equivalent to a Rabi oscillation exactly at resonance, for which
an exact solution is available for any function $q(\tau)$. For a
pulse extending from $\tau_1$ to $\tau_2$, the Rabi phase
$\varphi_r$ at the end of the pulse is given by :
\begin{equation}
\label{PR5} \varphi_r = \int_{\tau_1}^{\tau_2} (q^2/2) d\tau
\end{equation}
and if $\left|\psi(\tau_1)\right> =\left|\pm2\right>$, the final
state is :
\begin{eqnarray}
\label{PR6} \left|\psi(\tau_2)\right> & = & e^{\left[-i\left(
4(\tau_2 -\tau_1) +(\varphi_r/3)\right) \right]} \nonumber
\\ & \times & \left[ \cos\left( \frac{\varphi_r}{2} \right) \left|\pm2\right> -i
\sin\left(\frac{\varphi_r}{2}\right) \left|\mp2\right> \right]
\end{eqnarray} where the phase shift of the $|\pm2>$ states due to
their{\it mean} energy shift has been expressed as a fraction of
the Rabi phase. When $\left|\psi(\tau_1)\right> = \left|0\right>$,
the final state is the $\left|0\right>$ state with an extra phase
shift, also due to its energy shift:
\begin{equation}
\label{PR7} \left|\psi(\tau_2)\right>  = e^{i \varphi_r}
\left|0\right>
\end{equation} From now on, we consider a $\varphi_r =\pi$ pulse. If
the wavefunction at time $\tau_1$ is given by :
\begin{equation}
\label{PR8} \left|\psi(\tau_1)\right> = \sum_{p =-2,0,+2}
a_p(\tau_1) \left|p\right>
\end{equation}
the wavefunction at time $\tau_2$ is given by :
\begin{eqnarray}
\label{PR9} \left|\psi(\tau_2)\right> & = & e^{i\pi} a_0(\tau_1)
\left|0\right> + e^{\left[-4i(\tau_2 -\tau_1)
-(5i\pi/6)\right]}\nonumber
\\ & \times & \left[a_{-2}(\tau_1) \left|+2\right> + a_{+2}(\tau_1)
\left|-2\right>\right]
\end{eqnarray} The phase factor $\exp\left[-4i(\tau_2 -\tau_1)\right]$
is due to the free propagation of the $|\pm2>$ states and is not
linked to the diffraction process. The interesting results are the
diffraction phases equal to $(+\pi)$ for the $\left|0\right>$
state and $(-5\pi/6)$ for the $\left|\pm2\right>$ states. The
opposite signs of the diffraction phases are a consequence of the
opposite signs of the energy shifts of these levels. In the
resulting phase difference, the level shift contribution, equal to
$4\pi/3$, is proportional to the Rabi phase $\varphi_r$, taken
equal to $\pi$. In an experiment, this phase difference may differ
from this calculated value, as a result of an imperfect $\pi$
pulse or of other effects neglected here (e. g. : $\kappa \neq
0$).

\section{Simple model of the contrast interferometer of S. Gupta
et al}

\begin{figure}[htb]
\includegraphics[width=7 cm]{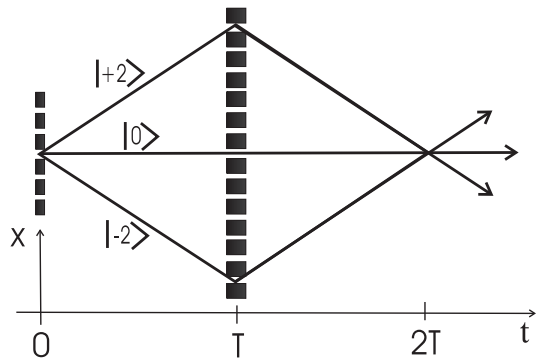}
\caption{\label{interferometer} In the $x$, $t$ plane, we have
represented the atomic paths followed by the wavepackets in the
interferometer of Gupta et al. [13] : Raman-Nath diffraction at
time $t=0$, second order Bragg diffraction at time $t=T$,
detection near time $t=2T$. }
\end{figure}

We now calculate the output signal of the contrast interferometer
developed by S. Gupta et al. \cite{gupta02}. This interferometer
uses second order Bragg diffraction and Raman-Nath diffraction and
the atomic paths are represented in figure~\ref{interferometer}.
The initial state is a Bose Einstein condensate, approximated here
by a $\left|\kappa=0\right>$ state. A first intense and brief
pulse from $\tau=0$ till $\tau_{RN}$ is used to diffract this
initial state in three coherent states, $\left|0\right>$,
$\left|\pm2\right>$. Within the Raman-Nath approximation, the
wavefunction for $\tau_{RN}$ is given by :
\begin{equation}
\label{CI1} \left|\psi(\tau_{RN})\right> = J_0 \left|0\right>  - i
J_1 \left[\left|+2\right> + \left|-2\right>\right]
\end{equation} the argument $\gamma$ of Bessel functions being
omitted for compactness. The best contrast \cite{gupta02} would be
obtained with diffraction probabilities equal to $50\%$ for the
$\left|0\right>$ state and $25\%$ for each of the $\left|\pm
2\right>$ states. It is impossible to fulfill perfectly these two
conditions simultaneously as the first one implies $\gamma = 1.13$
whereas the second one implies $ \gamma=1.21$. We can nevertheless
suppose that $\gamma \approx 1.17$. Although $J_2(1.17) \approx
0.15$, we will neglect here the second order diffraction
amplitudes, as done in reference \cite{gupta02}. We assume that
$\tau_{RN}$ is negligible so that free propagation starts at
$\tau=0$ and lasts till the Bragg diffraction pulse which extends
from $\tau_1$ to $\tau_2$. Using equation (\ref{PR9}), we get the
wavefunction after this pulse :
\begin{eqnarray}
\label{CI2} \left|\psi(\tau_2)\right> & = &  J_0 e^{i\pi}
\left|0\right> \nonumber
\\ & + & J_1 e^{-4i\tau_2} e^{-4i\pi/3} \left[\left|+2\right> + \left|-2\right>\right]
\end{eqnarray} Free propagation goes on till a time $\tau$
where the matter grating formed by the interference of these three
states is read by the reflection of a laser beam. The atomic
density as a function of $X$ and $\tau$ is deduced from the
wavefunction :

\begin{eqnarray}
\label{CI3} |\left<X|\psi(\tau)\right>|^2  & = & J_0^2
 + 2 J_1^2 ( 1 + \cos(4X) ) + 4 J_0J_1 \nonumber \\ & \times  &
 \cos(2X) \cos\left(4\tau +\frac{7\pi}{3}\right)
\end{eqnarray}
The experimental signal $S(\tau)$ is the intensity of the light
reflected by this grating. This homodyne detection signal is
proportional to the square of the $\cos(2X)$ modulation of the
atomic density, with the following time-dependence:
\begin{equation}
\label{CI4} S(\tau) \propto \cos^2\left(4\tau
+\frac{7\pi}{3}\right)
\end{equation}
while the equation used by Gupta et al. is :
\begin{equation}
\label{CI5} S(\tau) \propto \sin^2\left( 4 \tau \right)
\end{equation}
\noindent The difference between equation (\ref{CI4}) and
(\ref{CI5}) will be important only if one wants to make an
absolute prediction of the phase, but it has no consequence in the
analysis carried by S. Gupta et al. \cite{gupta02}, because their
fitted value of $\omega_{rec}$ comes from the derivative of the
phase with the time interval $T$ \cite{pritchard02pc}. However,
our result remains interesting as it may explain a large part of
the observed phase noise, $200$ mrad from shot to shot. In the $7
\pi/3$ phase of equation (\ref{CI4}), $4\pi/3$ are proportional to
the Rabi phase, which is itself proportionnal to $q^2$ i.e. to the
square of the laser power density during the Bragg pulse.
Therefore, a $1$\% variation of the laser power density changes
the diffraction phase by $84$ mrad.

Our calculation relies on several approximations, some of them
being not very accurate in the experimental conditions of  S.
Gupta et al. \cite{gupta02}:

i) the $\kappa =0$ approximation is an oversimplification but the
calculation with $\kappa\neq0$ is more complex.

ii) the first diffraction pulse used in the experiment is 1 $\mu$s
long, corresponding to  $\tau_{RN} = 0.157$. Assuming $\gamma
\approx 1.17$, we get $q \approx 3.7$ and the validity condition
(\ref{RN1}) requires $\tau \leq 0.13$. Therefore, the corrections
to the Raman-Nath phases are not fully negligible. We have also
neglected the second order diffraction beams, which contribute to
the signal.

iii) as for the perturbation expansion used to describe Bragg
diffraction, the $\pi$-pulse used is Gaussian with a width of $7.6
\mu$s \cite{gupta02}. Assuming that $q =q_{max}
\exp[-(t-T)^2/(2\sigma_t^2)]$, with $\sigma_t =3.8 \mu$s, i.e.
$\sigma_{\tau} \approx 0.6$, we get the value $q_{max} \approx
2.4$, well outside the validity range of our second order
perturbation expansion. Higher order terms in $q^n$ with $n=4,
6,..$ contribute to the phases and the sensitivity of the
diffraction phase to the laser power density may even be larger
than predicted above.

Obviously, to describe very accurately this experiment, a full
numerical modelization is needed and feasible, as the problem
reduces to a $1D$ Schr\"odinger equation, if atom-atom
interactions are neglected. But, as noted by Gupta et al., the
mean field effect of the condensate can also modify atomic
propagation and this effect has not been not considered here.

\section{Conclusion}

In this paper, we have made a simple and tutorial calculation of
the phase shifts of atomic waves due to elastic diffraction
process by a laser standing wave. We have calculated the
associated phase shift for the contrast interferometer of D.
Pritchard et al. \cite{gupta02}, thus showing that it should be
possible to make an  experimental test of the dependence of the
diffraction phase shifts with potential strength and interaction
time. The present calculations are simple because of our
assumptions : Raman-Nath limit or perturbative regime, vanishing
initial momentum $\kappa=0$. An accurate modelization of a real
experiment requires numerical integration of Schr\"odinger
equation to describe the diffraction dynamics without any
approximation.

We have considered only first and second order diffraction. Higher
diffraction orders up to order $8$, have been observed
\cite{giltner95a,giltner95b,koolen02} with moderate laser power
densities. The leading term of the coupling matrix element
responsible for diffraction order $n$ behaves like $q^n$
\cite{giltner95a} whereas the leading terms of the energy shifts,
responsible for the diffraction phase shifts, are always in $q^2$.
Therefore, for diffraction orders $n>2$, the control of the phase
shifts will require a full knowledge of the pulse shape. For the
second order of diffraction, the diffraction phase shifts and the
Rabi phase are simply related, as long as second order
perturbation theory is a good approximation.

We have made a systematic use of atomic Bloch states to describe
atom diffraction by laser, following our previous paper
\cite{champenois01}. The introduction of Bloch states to describe
atoms in a laser standing waves is due to Letokhov and Minogin
\cite{letokhov78,letokhov81} in 1978 and also to Castin and
Dalibard \cite{castin91} in 1991. Their use is rapidly expanding,
in particular to treat Bose-Einstein condensates in an optical
lattice, as reviewed by Rolston and Phillips \cite{rolston02}.
When coupled to reduced units as done here, the atomic Bloch
states represent a very efficient tool to get a simple
understanding of the diffraction process.

\vspace{1cm}

\section{Acknowledgements}

We thank C. Cohen-Tannoudji, J. Dalibard and C. Salomon for very
fruitfull discussions, D. Pritchard for a very useful private
communication and R\'egion Midi Pyr\'en\'ees for financial
support.



\end{document}